\documentclass[pra,a4paper,nofootinbib,twocolumn,showpacs,preprintnumbers,amsmath,amssymb,floatfix,amstex,superscriptaddress]{revtex4}
\usepackage{graphicx,graphics,rotating}         % Include figure files
\usepackage{dcolumn}                            % Align table columns on decimal point
\usepackage{bm,fancybox}                       % bold math
\usepackage{times,euscript,eufrak,oldgerm}              %
\usepackage[german,english]{babel}          %
\usepackage{psfrag}
\usepackage{color}
\newcommand{\ket}[1]{\vert#1\rangle}                    %
\newcommand{\bra}[1]{\langle #1\vert}                    %
\newcommand{\proj}[1]{\vert#1\rangle\langle#1\vert}
\newcommand{\abs}[1]{\vert#1\vert}
\DeclareMathOperator{\Tr}{Tr}
\DeclareMathOperator{\Neg}{Neg}

\begin{document}
\title{Scalability  of GHZ and random-state entanglement in the presence of decoherence} %
\author{Leandro Aolita}
\affiliation{ICFO-Institut de Ci\`encies Fot\`oniques, Mediterranean
Technology Park, 08860 Castelldefels (Barcelona), Spain}
\author{Daniel Cavalcanti}
\affiliation{ICFO-Institut de Ci\`encies Fot\`oniques, Mediterranean
Technology Park, 08860 Castelldefels (Barcelona), Spain}
\author{Antonio Ac\'in}
\affiliation{ICFO-Institut de Ci\`encies Fot\`oniques, Mediterranean
Technology Park, 08860 Castelldefels (Barcelona), Spain}
\affiliation{ICREA-Instituci\'o Catalana de Recerca i Estudis
Avan\c cats, Lluis Companys 23, 08010 Barcelona, Spain}
\author{Alejo Salles}
\affiliation{%
Instituto de F\'\i sica, Universidade Federal do Rio de Janeiro, Caixa Postal 68528, 21941-972 Rio de Janeiro, RJ, Brazil}
\affiliation{Physikalisches Institut der Albert-Ludwigs-Universit\"at, Hermann-Herder-Str.\ 3, D-79104 Freiburg, Germany}
\author{Markus Tiersch}
\affiliation{Physikalisches Institut der Albert-Ludwigs-Universit\"at, Hermann-Herder-Str.\ 3, D-79104 Freiburg, Germany}
\author{Andreas Buchleitner}
\affiliation{Physikalisches Institut der Albert-Ludwigs-Universit\"at, Hermann-Herder-Str.\ 3, D-79104 Freiburg, Germany}
\author{Fernando de Melo}
\affiliation{Physikalisches Institut der Albert-Ludwigs-Universit\"at, Hermann-Herder-Str.\ 3, D-79104 Freiburg, Germany}
\date{\today}%
%

%%
% Abstract
%%

\begin{abstract}
We derive analytical upper bounds
for the entanglement of generalized Greenberger-Horne-Zeilinger
states coupled to locally depolarizing and dephasing environments,
and for  local thermal baths of arbitrary temperature. These bounds
apply for {\it any} convex quantifier of entanglement, and
exponential entanglement decay with the number  of constituent particles is found.
The bounds are tight for depolarizing and dephasing channels. We also show that randomly generated initial states tend to violate these bounds, and that this discrepancy grows with the number of particles.
\end{abstract}
\maketitle
%%%%%%%%%%%%%%%%%%%%%%%%%%%%%%%%%%%%%%%%%%%%%%%%%%%%%%%%%%%%%%%%%%%%%%%%%%%%%%%%%%%%%%%%%%%%%%%%%%%%%%%%%%%%%%%%%%%%%%%%%%%%%%%%

%%%%%%%%%%%%%%%%%%%%%%%%%%%%%%%%%%%%%%%%%%%
%
% INTRODUCTION
%

\section{Introduction}

Impressive experimental progress in the manipulation
of composite quantum systems towards real-world applications of
quantum information and communication theory has taken place over the last few years.
The controlled production of genuinely
multipartite entangled states plays a central role in this
program, and is crucial for the scaling of existing toy quantum
protocols to mature technologies. Among those, the W~\cite{W} and
Greenberger--Horne--Zeilinger (GHZ)~\cite{GHZ} types of
entanglement play a  paradigmatic role, since they
incarnate characteristic traits and subtleties of multipartite
entanglement which allow, e.g., to implement protocols for secure
quantum communication.

Three photon W-type entanglement~\cite{Eibl, Witnesses}, and
five- and six- photon GHZ entangled states ~\cite{fiveGHZ,Lu} have
already been observed. Very recently, a ten party GHZ state was
produced using five hyperentangled photons~\cite{lastPan}. GHZ
states have also been reported in cavity QED
experiments~\cite{Raschenbeutel}, and for three~\cite{Roos},
four~\cite{Sackett} and six~\cite{Leibfried} trapped ions. Eight
ions were prepared in a W state~\cite{Haeffner}, and the
controlled generation of different multipartite entanglement
families was shown in~\cite{wieczorek}. Furthermore, the recent
implementation of a robust and extremely--high--fidelity
entangling gate of two ions~\cite{Jen} opens the way to the
controlled production of GHZ states of a few tens of ions.

However, it is known that scaling multipartite entangled states up
to many constituents is haunted by the decoherence processes arising
from the unavoidable and (most times) detrimental interaction of the system
degrees of freedom with the environment. Complete disentanglement
may even occur at finite
times~\cite{Zyczkowski,SimonKempe,DoddHalliwell,DuerBriegel,YuEberly,Franca,Terra,Terra2,Andre,fine,MintertPhysRep},
as experimentally demonstrated in~\cite{scienceESD, praESD,
Kimble}. It is therefore vital for experimental implementations to
predict the time scales on which appreciable amounts of
entanglement will prevail, under realistic assumptions on the
environment coupling. Theoretical studies of disentanglement
dynamics
for large systems were presented in~\cite{SimonKempe, DuerBriegel, Andre, fine, Hein, nacho07, Us, Guehne,Guehne2,Borras,MintertPhysRep},  
in particular for GHZ and W type states. While Refs.~\cite{SimonKempe,DuerBriegel,Hein} considered the scaling of the disentanglement time with the system size,
Refs.~\cite{Us,Andre,Guehne} put the focus on the scaling of the
short time scale behaviour of entanglement, what allows for a much better characterization of the robustness
of the initial state's entanglement under noise. In particular, it was shown in~\cite{Us}
that even if the disentanglement time of GHZ states grows with the number of parties $N$, the
residual entanglement (there quantified by the negativity~\cite{VidWer}) is reduced to
arbitrarily small values at
times which decrease with $N$.

In the present paper, we expand the above studies to larger
classes of multipartite entangled states. We first investigate the
scaling properties of disentanglement of mixed-state
generalizations of GHZ states, namely, generalized GHZ-diagonal states. For systems subject to natural decoherence models,
analytical upper bounds can be derived for their entanglement and for the
associated scaling behavior with the system size. Our findings are valid for {\em any} convex
entanglement measure. In addition, we numerically study the entanglement dynamics
of randomly generated states, quantified in terms of the
respective states' negativity. Random
samples of pure initial states do not abide to the above-mentioned bounds. Our
numerical data suggest that the latter discrepancy increases with
the number of system constituents, which gives evidence of the
exponential fragility of GHZ entanglement towards
decoherence {\em not} being a generic feature.

The paper is organized as follows. In Sec.~\ref{sec:States}, we introduce our notation and define the generalizations of the GHZ
states to be studied later. Sec.~\ref{sec:Noise}
defines our environment models. The analytical upper bounds for
the entanglement, together with the resulting
scaling behavior, are derived
in Sec.~\ref{sec:Bounds}, where the tightness of these bounds is also assessed. 
Sec.~\ref{sec:Random} compares the scaling of the properties of GHZ-entanglement with those of random pure states. Finally, Sec. \ref{Conclu} summarizes our conclusions.

%%%%%%%%%%%%%%%%%%%%%%%%%%%%%%%%%%%%%%%%%%%
%
% STATES
%

\section{Generalized GHZ and GHZ diagonal states}
\label{sec:States}

Originally, the GHZ-state~\cite{GHZ} was defined for three qubits as a superposition of the many particle states of all of them being in their respective basis state $\ket{0}$ and all of them being in the orthogonal state $\ket{1}$. This notion straightforwardly extends to more qubits and gives the GHZ-state to be
\begin{equation}
\ket{\psi_\text{GHZ}} = \frac{1}{\sqrt{2}} \big( \ket{000\ldots 0} + \ket{111\ldots 1} \big)
\,.
\end{equation}
Without imposing more details on the nature of the
specific quantum system the basis states $\ket{0}$ and $\ket{1}$ are abstract and without any physical meaning. Thus, formally, the state $(\ket{010}+\ket{101})/\sqrt{2}$ for three qubits,
e.g., follows the spirit of the GHZ
construction in precisely the same manner. When considering all
such states it is helpful to interpret a string of zeros and ones as
the binary representation of a number, the number of the second term always being
bit-wise inverted with respect to the first.
A many particle basis state of $N$ qubits can thus be labeled with numbers from 0 (all zeros) to $2^N-1$ (all ones). Our three qubit example would therewith read $(\ket{2}+\ket{5})/\sqrt{2}$.
If we furthermore allow for different amplitudes of the vectors and, for later convenience, treat a phase difference of $\pi$ separately (thus distinguishing
states of even and odd parity), we arrive at the 
{\em generalized GHZ states} of $N$ qubits,
\begin{equation}
\label{kalphabeta}
\ket{\psi_k^{\pm}(\alpha,\beta)}\equiv\alpha\ket{k}\pm\beta\ket{\bar{k}},
\end{equation}
with an $N$-bit number $k$, $0\le k\le 2^N-1$, $\bar{k}$
the bit-wise flipped number, and complex amplitudes $\alpha$ and $\beta$ such that $\abs{\alpha}^2+\abs{\beta}^2=1$, and $\alpha,\beta \neq 0$.

An incoherent mixture of several generalized GHZ-states $\ket{\psi^\pm_k(\alpha,\beta)}$ with different $k$ and
parity, but the same amplitudes $\alpha$ and $\beta$, is a {\em generalized GHZ diagonal state}~\cite{Duer&Cirac}:
\begin{multline}
\label{GGHZDrho0}
\rho=\sum_{k=0}^{2^{N}-1} \Big( \lambda_k^+ \proj{\psi_k^{+}(\alpha,\beta)}+\\
+\lambda_k^-\proj{\psi_k^{-}(\alpha,\beta)} \Big)
\,.
\end{multline}
The coefficients $\lambda^\pm_k$ denote the respective
probabilities with which the state appears in the mixture.
Naturally, they are positive and sum up to one.

%%%%%%%%%%%%%%%%%%%%%%%%%%%%%%%%%%%%%%%%%%%
%
% ENVIRONMENTS
%

\section{Noise models}
\label{sec:Noise}

Given the above class of initial states, we proceed now to describe three paradigmatic models of incoherent dynamics present in typical experimental settings. We assume that qubits do not interact with
each other, neither directly, nor indirectly through their baths,
i.e., they feel independent and local environments. This is a reasonable
approach for qubits being located in different laboratories, or
sufficiently well isolated from each other.
Furthermore, we assume that the qubit-bath interaction is
identical for all of them.

We describe the state evolution by means of a map (or channel) such that a single qubit's initial state $\rho_i$ is mapped onto
its final state by virtue of $\rho_f=\mathcal{E}(\rho_i)$. Since the mapping is between quantum states, it needs to be completely positive
and trace preserving. Every such map allows for an operator sum (or Kraus) representation~\cite{NielsenChuang}
\begin{equation}
\mathcal{E}(\rho_i) = \sum_j K_j \rho_i K_j^\dag\, ,
\end{equation}
with Kraus operators $K_j$ fulfilling $\sum_j K_j^\dag K_j=\openone$.
The initial joint state $\rho$ of $N$ qubits evolves thus according to the $N$-fold tensor product of the individual maps:
\begin{equation}
\label{composition}
\rho\equiv \Lambda(\rho)
\equiv
\underbrace{{\cal E}\otimes{\cal E}\otimes \dots \otimes{\cal E}}_N (\rho)
\,.
\end{equation}

Specifically, the channels we consider~\cite{NielsenChuang} are: the depolarizing channel (that
occurs, e.g., in spin scattering), the dephasing (or phase
damping) channel (typical of elastic collisional interactions),
and a thermal bath at arbitrary temperature (provided, e.g., by a
thermal radiation background). 

The depolarizing channel (for which
we will use the subscript D in the sequel) describes the situation
in which the environment isotropically destroys the information on
a qubit's state and thus steers it into a maximally mixed state
$\openone/2$. The characteristic quantity describing the dynamics
is the probability $p$ of finding the state completely
depolarized. The single qubit's initial state $\rho_i$ thus
evolves to ${\cal E}_\text{D}(\rho_i)=\rho_i(1-p)+{\openone}p/2$.
The corresponding Kraus representation allows for a convenient
form in terms of the Pauli matrices:
\begin{equation}
\label{DCPauli}
 {\cal E}_\text{D}(\rho_i)=\sum_{j=0}^{3}s_j\sigma_j\rho_i\sigma_j
 \,,
\end{equation}
where $s_0\equiv 1-3p/4$, $s_1=s_2=s_3\equiv p/4$, $\sigma_0\equiv \openone$, and $\sigma_1$, $\sigma_2$, and $\sigma_3$ are the three familiar Pauli operators associated with the respective qubit.

As a second type of noise, we discuss the phase damping channel (subscript PD). It represents
a situation when quantum coherence is lost
{\em without} any population or excitation exchange. If the probability of complete phase loss is $p$, a Kraus
representation reads
\begin{multline}
\label{PD}
{\cal E}_\text{PD}(\rho_i)=(1-p)\rho_i+ \\
+ p\Big(\proj{0}\rho_i\proj{0}+\proj{1}\rho_i\proj{1}\Big)
\end{multline}
with Kraus operators
\begin{equation} \label{eq:KrausPD}
M_0 = \openone\sqrt{1-p} \,,
M_1 = \proj{0}\sqrt{p} \,,
M_2 = \proj{1}\sqrt{p} \,.
\end{equation}

Finally, the third type of environment to be dealt with here
is a thermal bath (subscript T).
In this scenario, the qubit's basis states $\ket{0}$ and $\ket{1}$ function as ground and excited state, respectively, in order to exchange excitations with the bath. In the Born--Markov approximation~\cite{Preskill,Cohen} this yields the Kraus operators
\begin{align} \label{eq:KrausT}
K_0 & = \sqrt{\frac{\bar{n}+1}{2\bar{n}+1}}\Big(\proj{0}+\proj{1}\sqrt{1-p}\Big) \,, \nonumber \\
K_1 & = \sqrt{\frac{\bar{n}+1}{2\bar{n}+1}}\ket{0}\bra{1} \sqrt{p}\,, \\
K_2 & = \sqrt{\frac{\bar{n}}{2\bar{n}+1}}\Big(\proj{0}\sqrt{1-p}+\proj{1}\Big) \,, \quad \text{and} \nonumber \\
K_3 & = \sqrt{\frac{\bar{n}}{2\bar{n}+1}}\ket{1}\bra{0}\sqrt{p} \nonumber
\,,
\end{align}
where the average excitation $\bar{n}$ of the bath modes induces
an energy exchange from the bath to the system, and vice versa. For zero temperature, i.e., $\bar{n}=0$, this reduces to the amplitude damping
channel, where system excitations irreversibly dissipate into the bath, triggered by the vacuum fluctuations of the bath mode, with a rate $\gamma$.
The opposite case, at infinite temperature, is established by taking the limit
$\bar{n}\rightarrow\infty$ and $\gamma\rightarrow0$, with
$\bar{n}\gamma=\Gamma\equiv {\rm const}$, and
models a purely diffusive environment with diffusion constant $\Gamma$. In this case, $p$ is the probability for an excitation exchange between system and bath.

We stress that in all three cases, the models presented can
encompass many different dynamics, depending on how one relates
$p$ to time $t$. For instance, the thermal bath at zero
temperature can model an atom interacting with the free
electromagnetic field if we take $p = 1-\exp(-\gamma t/2)$, or with the field in a cavity by taking $p=\sin^2(\omega
t/2)$, in which Rabi oscillations (of vacuum Rabi frequency $\omega$)
can take place.

%%%%%%%%%%%%%%%%%%%%%%%%%%%%%%%%%%%%%%%%%%%%%%%%%%%%%%%%%%%%%%%%%%%%%%%%%%%%%%%
%
% BOUNDS
%

\section{Bounds on the robustness of entanglement}
\label{sec:Bounds}

With the detailed description of the states and the dynamics to be
scrutinized at hand we can assess the robustness of the system's
entanglement by bounding it from above as it evolves according to
such open dynamics. We only assume two basic properties of the entanglement quantifier
$E$ and thus provide results for \emph{any} such entanglement
quantifier. 
We merely require $E$ to vanish for separable states, and not to increase when probabilistically
mixing two states $\sigma$ and $\omega$, i.e., it needs to be
convex:
\begin{equation} \label{eq:convexity}
E \big[ \mu\sigma+(1-\mu)\omega \big]
\leq
\mu E(\sigma)+(1-\mu)E(\omega)
\end{equation}
for any $\mu\in[0,1]$.

The basic idea behind the derivation of the upper bounds 
is to decompose all studied states as  convex
combinations of an entangled and a separable part, i.e., 
$\rho=\mu_{ent}\rho_{\text{ent}}+(1-\mu_{\text{ent}})\rho_{\text{sep}}$, and to use the properties
listed just above to bound the entanglement evolution as $E(\rho)\leq\mu_{\text{ent}}
E(\rho_{\text{ent}})$. 

In what follows we identify such decompositions for initial states of
the general GHZ or general GHZ diagonal type, under the influence
 of the paradigmatic environments considered in the previous section.

%++++++++++++++++++++++++++++++++++++++++++++++++++++++++++++++++++++++++++++++
\subsection{Depolarization}
\label{Depol}
%++++++++++++++++++++++++++++++++++++++++++++++++++++++++++++++++++++++++++++++

We begin with the case of open system dynamics that independently
depolarizes each qubit. For this type of decoherence process, we
prove now the following:

\begin{itemize}
\item[{\it (i)}] The entanglement of an initially generalized GHZ diagonal state $\rho$ subject to local depolarizing $\Lambda_\text{D}$ is bounded from above as
\begin{equation}
\nonumber
E \left( \Lambda_\text{D}(\rho) \right) \le (1-p)^N E(\rho).
\end{equation}
\end{itemize}

To prove this statement, let us first begin by considering a single
generalized GHZ state, namely the particular case
$\rho_0\equiv\proj{\psi_0^{+}(\alpha,\beta)}$.

The evolved density matrix $\Lambda_\text{D}(\rho_0)$ is
straightforwardly obtained by the $N$-fold application of the
depolarizing channel~\eqref{DCPauli} of one qubit according to
Eq.~\eqref{composition}. In the resulting $N$-fold operator sum we
first focus on one of the sums for a single qubit. There the two
terms for $\sigma_1$ and $\sigma_2$ yield contributions of equal
weight ($s_1=s_2$) with different signs in their coherences,
because $\sigma_1$ does a bit-flip whereas $\sigma_2$ results in a
combined bit- and phase-flip (parity change). Altogether, this
cancels the coherences and thus results in a diagonal and hence
separable contribution. Thus, again considering all the sums, the
only terms not immediately causing separable contributions are the
applications of only the identity ($\sigma_0$) or the parity
changing  operator $\sigma_3$.

Therefore, after the application of channel
$\Lambda_\text{D}$, we identify three contributions to the final
state. First, there is the unchanged state
$\proj{\psi_0^{+}(\alpha,\beta)}$ resulting from the application
of only identities or an even number of parity changes. Second,
there is the parity changed counterpart
$\proj{\psi_0^{-}(\alpha,\beta)}$ that is generated by an odd
number of parity changes. Lastly, there is a separable
contribution originating from the application of at least one
$\sigma_1$ or $\sigma_2$ to one of the qubits:
\begin{align}
\Lambda_\text{D} \big( \rho_0 \big) %\proj{\psi_{0}^{+}(\alpha,\beta)}\big)
=&
\lambda_+ \proj{\psi_{0}^{+}(\alpha,\beta)} \nonumber \\
+&\lambda_- \proj{\psi_{0}^{-}(\alpha,\beta)} \nonumber \\
+&\lambda_\text{sep} \rho_\text{sep}
\,.
\end{align}
Here, the first two terms are of opposite parity. Their
coherences partially cancel when being summed,  yielding another
diagonal (separable) contribution. Their difference is what
determines the remaining entangled contribution in the
decomposition. Its careful evaluation (see
Appendix~\ref{app:Details(i)}) yields
\begin{equation}
\Lambda_\text{D} \big( \rho_0 \big) %\proj{\psi_{0}^{+}(\alpha,\beta)}\big)
=
(1-p)^N \rho_0  %\proj{\psi_{0}^{+}(\alpha,\beta)} \nonumber \\
+ \left[1-(1-p)^N\right] \rho^\prime_\text{sep}
\,.
\end{equation}
This result does not depend on the fact that it was derived using $\ket{\psi_{k}^{+}(\alpha,\beta)}$ with $k=0$ as initial state and holds for different $k$ as well.
Also, since the application of channels is linear this result extends immediately to any convex combination such as the generalized GHZ diagonal states given in Eq.~\eqref{GGHZDrho0}. The convexity property~\eqref{eq:convexity} then yields the desired result.

As a corollary of {\it (i)}, we have:
\begin{itemize}
\item[{\it (ii)}] For the special case of two qubits ($N=2$), the bound in {\it (i)} holds for \emph{all} initial two-qubit states (not only the general GHZ diagonal ones).
\end{itemize}

Consider first the pure-state case. Any pure two-qubit state $\ket{\Psi}$ can be expressed in local product-state bases such that it is
$\ket{\Psi} =
\alpha\ket{00}+\beta\ket{11}\equiv\ket{\psi_{0}^{+}(\alpha,\beta)}$,
with real and positive $\alpha$ and $\beta$~\cite{NielsenChuang}.
Then, for {\it any} pure two-qubit state, there is always a local
basis (the Schmidt basis) in which it is  a generalized GHZ state.
Now, since ${\cal E}_\text{D}$ is basis-independent -- it shrinks
the Bloch sphere without distinction of any particular direction  --, the local basis to which all four Kraus
operators in \eqref{DCPauli} make reference can be any, in
particular that in which $\ket{\Psi}$ is a generalized GHZ state.
By employing {\it (i)}, we have that {\it (ii)} holds for {\it
all} pure two-qubit states.

For the mixed state case, we consider {\it any} two-qubit state $\rho$ in its {\it convex roof} \cite{Bennett} decomposition $\rho \equiv \sum_{n} {p_n}\proj{{\Psi_n}}$ (for which the entanglement and the average entanglement coincide) and repeat the same reasonings used before.
Linearity implies that $\Lambda_\text{D}(\rho) = \sum_{n}{p_n}\Lambda_\text{D}(\proj{{\Psi_n}})$.
Convexity implies that $E[\Lambda_\text{D}(\rho)]\le\sum_{n}{p_n}E\big[\Lambda_\text{D}(\proj{{\Psi_n}})\big]$, with the latter being the average entanglement of $\Lambda_\text{D}(\rho)$. The latter is in turn smaller or equal than $\sum_{n}{p_n} (1-p)^2 E(\ket{{\Psi_n}})$, for {\it (i)} holds for any pure two-qubit state, so in particular also for all of the $\ket{{\Psi_n}}$. Therefore, we have:
\begin{eqnarray}
\label{iii}
\nonumber
E\big(\Lambda_\text{D}(\rho)\big)&\le&\sum_{n}{p_n}(1-p)^2 E(\ket{{\Psi_n}})\\
\nonumber
&=&(1-p)^2 E\big(\sum_{n}{p_n}\ket{{\Psi_n}}\big)\\
&=&(1-p)^2 E(\rho),
\end{eqnarray}
where from the first to the second line of \eqref{iii} we used the
equivalence between the average entanglement and the entanglement
itself for the optimal decomposition.

It is important to mention that, as was anticipated in the
introduction, bound {\it (ii)} -- mathematically expressed in
\eqref{iii} -- is directly connected  to a universal law
discovered in \cite{konrad}. There, a universal bound on
$E[\Lambda(\rho)]$ was set for the particular case of $E$ being
the concurrence, but for ${\cal E}$ any completely-positive map.
Bound {\it (ii)} thus generalizes \cite{konrad} in terms of the allowed entanglement quantifier $E$. It restricts, however, the environment coupling to the particular case of depolarization dynamics and yields a slightly weaker bound when evaluated for $E$ being the concurrence.

We also stress that bounds {\it (i)} and {\it (ii)} are optimal
with respect to the class of states and entanglement quantifiers
we deal with.
This can be seen directly
by showing an entanglement quantifier saturating the bound for at
least one state, which was done already in Ref.~\cite{Us}, where
it was shown that the most resistant negativities of evolved
GHZ-states tend to $(1-p)^N$  times their initial value in the
small $p$ or large $N$ limits.

%++++++++++++++++++++++++++++++++++++++++++++++++++++++++++++++++++++++++++++++
\subsection{Dephasing}
\label{Dephas}
%++++++++++++++++++++++++++++++++++++++++++++++++++++++++++++++++++++++++++++++

For individual dephasing environments the results are similar to the previously obtained. In this case we prove the following:

\begin{itemize}
\item[{\it (iii)}]  The entanglement of an initially generalized GHZ diagonal state $\rho$ subject to local dephasing $\Lambda_\text{PD}$ is bounded from above as
\begin{equation}
E\big(\Lambda_\text{PD}(\rho)\big) \le (1-p)^{N} E(\rho)
\,.
\end{equation}
\end{itemize}

The strategy to show this is the same as in \textit{(i)}, but
now with the dephasing channel~\eqref{PD}. In this fashion, we
first identify the terms in $\Lambda_\text{PD}(\rho_0)$ for a
single general GHZ state $\rho_0=\proj{\psi_0^{+}(\alpha,\beta)}$
that sum up to some separable state. We notice then that only the
first Kraus operator of ${\cal E}_\text{PD}$ (proportional to the
identity operator) yields a non(-necessarily)-separable state upon
application on $\ket{\psi_{0}^{+}(\alpha,\beta)}$. All other terms
contain at least one diagonal projector, $\proj{0}$ or $\proj{1}$,
that eliminates the coherences of $\ket{\psi_0^{+}(\alpha,\beta)}$
and takes it to some diagonal and hence separable matrix.
Therefore, a decomposition of the final state is
\begin{eqnarray}
\label{finalPD}
\Lambda_\text{PD} (\rho_0) = (1-p)^{N} \rho_0 + [1-(1-p)^N] \rho_\text{sep},
\end{eqnarray}
where again $\rho_{sep}$ is some fully separable density operator;
and where the procedure just used can be applied once more to any
generalized GHZ-state $\ket{\psi_k^{+}(\alpha,\beta)}$. The rest
of the proof follows thus as in \textit{(i)} from the linearity of
the channel and the convexity property of the entanglement
quantifier.

We should note that, in contrast to the depolarizing channel, the
dephasing channel is indeed basis-dependent -- it does not commute
with all local unitaries. Even though any pure two-qubit state is
of generalized GHZ form in its Schmidt basis, this basis is not
necessarily the one in which the dephasing channel is defined.
Thus {\it (iii)} does not generalize to all two-qubit states as
was the case with {\it (i)}. We also stress that again bound {\it
(iii)} is tight, as follows from a similar argument as for bounds
{\it (i)} and {\it (ii)}.

%+++++++++++++++++++++++++++++++++++++++++++++++++++++++++++++++++++++++
\subsection{A thermal bath}
\label{Therm}
%+++++++++++++++++++++++++++++++++++++++++++++++++++++++++++++++++++++++

For the last  example of environment models
treated in this work we prove the following:
\begin{itemize}
\item[{\it (iv)}] The entanglement of any initially
generalized GHZ-state
$\rho_k=\proj{\psi_k^{\pm}(\alpha,\beta)}$ subject to local
thermal baths $\Lambda_\text{T}$ with an average excitation
$\bar{n}$ in the bath modes is bounded from above as
\begin{widetext}
\begin{eqnarray}
\nonumber
E\big[ \Lambda_\text{T}(\rho_k) \big]
\le
\left[ \abs{\alpha}^2 \left(1-\frac{\bar{n}}{2\bar{n}+1}p
\right)^{N-\kappa}
\left( 1-\frac{\bar{n}+1}{2\bar{n}+1}p \right)^{\kappa}
+
\abs{\beta}^2 \left( 1-\frac{\bar{n}}{2\bar{n}+1}p
\right)^{\kappa}
\left( 1-\frac{\bar{n}+1}{2\bar{n}+1}p \right)^{N-\kappa}
\right]
E_\text{max},
\end{eqnarray}
\end{widetext}
\end{itemize}
where $\kappa$ is the number of ones in the binary string
$k$, i.e.\ the number of excitations in the state
$\ket{k}$, and $E_\text{max}$ is the maximal value that the
entanglement quantifyer $E$ can take.

As before, we start by considering the
final state in the operator sum representation of the
channel acting on the initial state and identify terms that
cancel to separable contributions to the state. Similarly
to {\it (iii)} we find that only terms with the two Kraus
operators $K_0$ or $K_2$, out of all four of
\eqref{eq:KrausT}, applied to all of the qubits constitute
the contributions that render a non(-necessarily)-separable
state. That is,
$\Lambda_\text{T}(\rho_k)=\sum_{j_{1},\ldots
j_{N}=0,2}K_{j_1}\otimes\ldots\otimes K_{j_N}\rho_k
K^\dag_{j_1}\otimes\ldots\otimes
K^\dag_{j_N}+\lambda_\text{sep}\rho_\text{sep}$. For
brevity we refer with $\lambda_\text{ent}\rho_\text{ent}$
to the first, in general non-separable term.
By virtue of the entanglement quantifier's convexity the
final entanglement is bounded from above, as in the
previous cases, by the probability $\lambda_\text{ent}$ of
the state contribution $\rho_\text{ent}$ in the first term
in the convex sum, times its entanglement, and, trivially, by
the maximal entanglement:
\begin{equation}
E[\Lambda_\text{T}(\rho_k)] \leq \lambda_\text{ent}
E(\rho_\text{ent})
\leq \lambda_\text{ent} E_\text{max}
\,.
\end{equation}
The probability $\lambda_\text{ent}$ is given by the trace
of the first term.
It is
\begin{widetext}
\begin{eqnarray}
\nonumber
\lambda_\text{ent}&=&\Tr \left[ \sum_{j_1,\ldots j_N=0,2}
K^\dag_{j_1}K_{j_1}\otimes\cdots\otimes
K^{\dagger}_{j_{N}}K_{j_{N}}\rho_0\right]
\nonumber\\
&=&\Tr \left[ \prod_{i=1}^N
\left(\left(1-\frac{\bar{n}}{2\bar{n}+1}p\right)\proj{0}+\left(1-\frac{\bar{n}+1}{2\bar{n}+1}p\right)
\proj{1} \right) \rho_k \right] \\
\nonumber
&=&\bra{\psi_k^{\pm}(\alpha,\beta)} \prod_{i=1}^N
\left[\left(1-\frac{\bar{n}}{2\bar{n}+1}p\right)\proj{0}+\left(1-\frac{\bar{n}+1}{2\bar{n}+1}p\right)\proj{1}\right]
\ket{\psi_k^{\pm}(\alpha,\beta)}\\
\nonumber
&=&\left[\abs{\alpha}^2\left(1-\frac{\bar{n}}{2\bar{n}+1}p\right)^{N-\kappa}\left(1-\frac{\bar{n}+1}{2\bar{n}+1}p\right)^{\kappa}+
\abs{\beta}^2\left(1-\frac{\bar{n}}{2\bar{n}+1}p\right)^{\kappa}\left(1-\frac{\bar{n}+1}{2\bar{n}+1}p\right)^{N-\kappa}\right],
\end{eqnarray}
\end{widetext}
where the explicit definitions of $K_0$ and $K_2$ in
\eqref{eq:KrausT} were used in the second
equality.

The scaling factor in the bound of \textit{(iv)} includes $\kappa$
as well as the amplitudes $\alpha$ and $\beta$, and thus still
incorporates details of the initial states. In order to arrive at
a scaling factor that only includes details of the thermal
environment and, of course, the number of qubits, we maximize the
expression over all states $\ket{\psi^\pm_k(\alpha,\beta)}$ which
we are focusing on. The solution is reached, for instance, in the
limit $\abs{\alpha}^2 \to 1$ and $\kappa=0$. This allows us to get
a larger, but state independent, scaling factor such that we can
rephrase \textit{(iv)}:

\begin{itemize}
\item[{\it (v)}] The entanglement of any initially
generalized GHZ-state
$\rho_k=\proj{\psi_k^{\pm}(\alpha,\beta)}$ subject to local
thermal baths $\Lambda_\text{T}$ with an average excitation
$\bar{n}$ in the bath modes is bounded from above as
\begin{eqnarray}
E\big[ \Lambda_\text{T}(\rho_k) \big]
\le
\left( 1-\frac{\bar{n}}{2\bar{n}+1}p \right)^N
E_\text{max}.
\end{eqnarray}
\end{itemize}

It is worth stressing that for this kind of noise model we
are restricted to the (pure) generalized GHZ-states. Also,
notice that for different temperatures, and thus  different
mean numbers of bath excitations $\bar n$, bounds {\it (iv)}
and \textit{(v)} range from
$E\big[\Lambda_\text{T}(\rho_k)\big] \le
\big[\abs{\alpha}^2(1-p)^\kappa+\abs{\beta}^2(1-p)^{N-\kappa}\big]E_\text{max}$
for bound \textit{(iv)} and a trivial bound for
\textit{(v)}, $E\big[\Lambda_\text{T}(\rho_k)\big] \le
E_\text{max}$, in the purely-dissipative, zero-temperature
limit with $\bar{n}=0$ (amplitude damping channel), to
$E\big[\Lambda_\text{T}(\rho_k)\big] \le (1-\frac{p}{2})^N
E_\text{max}$ for both bounds in the purely-diffusive,
infinite-temperature limit $\bar{n}\to\infty$.

\begin{figure*}[h!tbp]
\begin{center}
\includegraphics[width=1\linewidth]{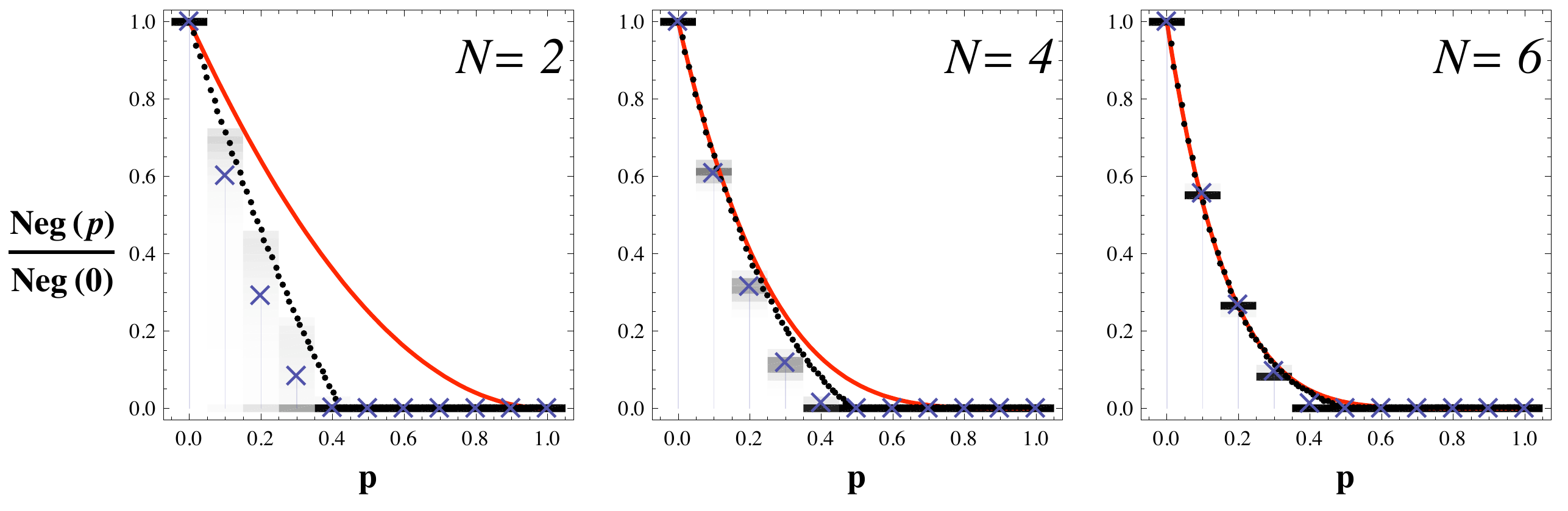}
\caption[width=2\linewidth]{ \label{Fig1} (Color online) Normalized negativities of most balanced bipartitions as a function of $p$ for systems of different size $N$ undergoing individual depolarization. The red solid line corresponds to the bound $(1-p)^N$, the dotted line to the balanced GHZ-state $\ket{\psi_0^{+}\big(1/\sqrt{2},1/\sqrt{2}\big)}$, which always lies below the bound $(1-p)^N$ and approaches it as $N$ grows, and the blue crosses to the average over 10000 initially random pure states distributed uniformly over the system Hilbert space. The gray shadings are the distributions of the normalized negativities around their mean values (bin population reflected by saturation, i.e.\ dark is the maximum and white the minimum). For $N=6$ the average normalized negativity no longer lies below the bound $(1-p)^N$. See text.}
\end{center}
\end{figure*}

\begin{figure*}[h!tbp]
\begin{center}
\includegraphics[width=1\linewidth]{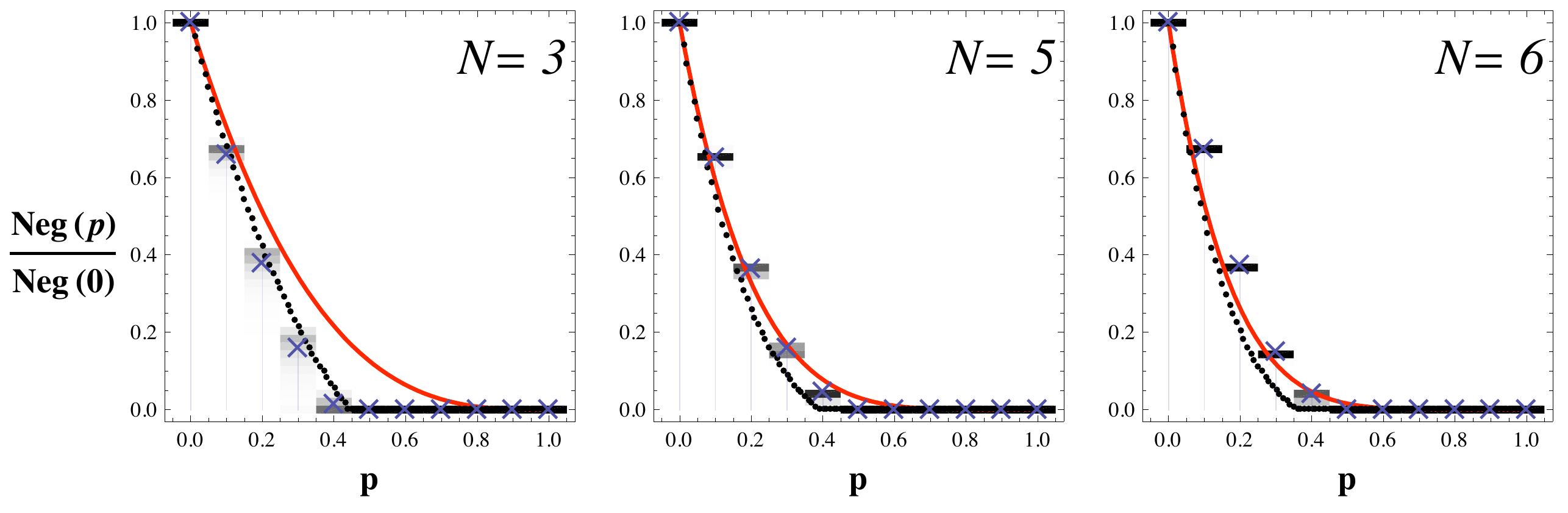}
\caption{\label{Fig2} (Color Online) Same as Fig. \ref{Fig1}, but for the least-balanced bipartitions. In this case the violation of the $(1-p)^N$ scaling law already appears at  $N=5$. See text.}
\end{center}
\end{figure*}

%%%%%%%%%%%%%%%%%%%%%%%%%%%%%%%%%%%%%%%%%%%%%%%%%%%%%%%%%%%%
%
% RANDOM
%
\section{Beyond the generalized GHZ states: comparison with random states}
\label{sec:Random}

The simplicity of the obtained bounds raises the question of how dependent 
they are on the specific choice of initial states. Some of the above 
bounds are in this sense more general than others, but all feature an 
 exponential scaling in the number $N$ of qubits, which renders their 
entanglement more fragile for increasingly many qubits.

In order to verify up to what extent the bounds are representative to a larger class of states than only to the ones they were actually derived for, we consider initial random pure states uniformly distributed over the entire Hilbert space~\cite{commentHaar}. As a simple, convex entanglement quantifier we choose negativity evaluated for a bipartition of the $N$ subsystems. For systems with varying number of constituents, we generate a sample of 10000 initial states, evolve it under the three considered incoherent dynamics, and calculate the negativity as a function of the probability $p$ of an incoherent event.

In general, we observe that such states violate the above bounds,
more drastically as we increase the number of parties $N$.
Figs.~\ref{Fig1} and~\ref{Fig2} show examples of this violation.
There, the average negativities $\Neg(p)$ (normalized to their initial value $\Neg(0)$) of a sample of 10000 initial pure states undergoing individual
depolarization is plotted for systems of different size $N$ (blue crosses). For comparison, they also contain the bound $(1-p)^N$ (red solid line)
and the specific case of the balanced GHZ state $\ket{\psi_0^{+}\big(1/\sqrt{2},1/\sqrt{2}\big)}$ (dotted line).
Necessarily, the latter always lies below the bound $(1-p)^N$ and approaches it as $N$ grows.
The average normalized negativity, however, violates the $(1-p)^N$ exponential-decay scaling law, in particular for larger $N$.
The gray shadings along the vertical direction represent the histogram of the samples' normalized negativities in gray-scale (bin population reflected by saturation, i.e.\ black the maximum and white the minimum)~\cite{commentgray}.
Fig.~\ref{Fig1} presents the data for negativity evaluated for the most balanced partition of $N/2$ versus $N/2$, where the violation is apparent for 6 qubits. Contrary, Fig.~\ref{Fig2} shows the least balanced partition of $1$ versus $N-1$
qubits, where a violation appears for 5 qubits.

\begin{figure}[htbp]
\begin{center}
\includegraphics[width=1\linewidth]{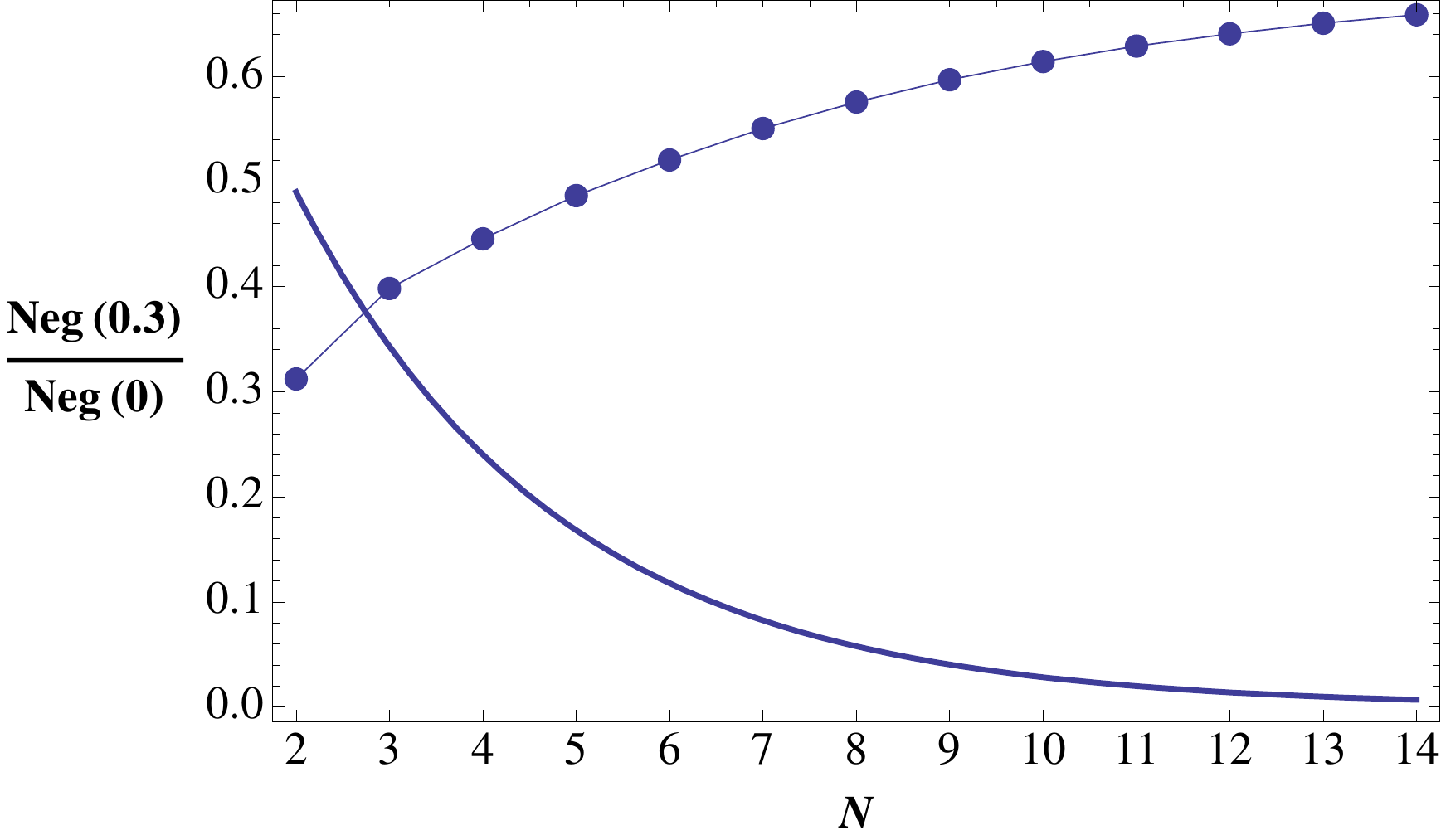}
\caption{(Color Online) Mean normalized negativity for the most unbalanced partition as a function of the system size, for a fixed value of the probability $p=0.3$, and random initial states undergoing the action of the dephasing channel (circles, joined by straight lines as a guide to the eye). Bound {\it (iii)} is plotted as a solid line.}
\label{Fig3}
\end{center}
\end{figure}

With the observation that initial random states violate the bounds, we can consider an extension of these bounds in such a fashion that these states are included as well without losing the scaling behaviour, i.e., an entanglement upper
bound that decreases with increasing $N$ for all the states.
Fig.~\ref{Fig3} suggests that this is \emph{not} possible -- the
mean value of the normalized negativity, of the least balanced
bipartition, at a given evolution step (dephasing at $p=0.3$)
grows with increasing $N$. In this plot, the full line represents
the the bound $(1-p)^N$, while the dots stand for the numerical
values obtained by a similar sampling as explained above.
However, due to the big computational effort of sampling many states of
large $N$, we rely on the negativity concentration effect for
large dimensional systems under incoherent dynamics~\cite{commentgray,concentration,hayden} in order to
pick just few typical states. The sample sizes are displayed in
Table~\ref{table}.

\begin{table}
\begin{tabular}{ccc}
  %\hline
  % after \\: \hline or \cline{col1-col2} \cline{col3-col4} ...
  N && Sample size \\
  \hline
  2 to 7 && 10000 \\
  8, 9, 10 && 5000 \\
  11 && 1560 \\
  12 && 100 \\
  13 && 10 \\
  14 && 1 \\
  %\hline
\end{tabular}
\caption{Sample size used in the numerical calculation of Fig.~\ref{Fig3}.}\label{table}
\end{table}

A persistently increasing mean negativity, which we confirmed numerically up to $N=14$ qubits, can thus not be bounded from above by \emph{any} nontrivially decreasing function. %Furthermore, this behaviour, which we only find for highly unbalanced partitions, is contrary to the general GHZ-states as it renders states more robust under dephasing. Therefore, 
This also implies that the exponential fragility of GHZ-type states is not typical, as it is not a decisive feature present in the ensemble of randomly chosen initial states.

The increasing entanglement robustness for random states is observed for all channels here scrutinized. However, it only takes place for highly unbalanced partitions, where the entanglement between few qubits with the remaining system is probed. For more balanced partitions, although the bounds presented in the previous section are violated (as shown in Fig.~\ref{Fig1}),  we observe a stronger entanglement decay of random states for increasing number of subsystems. Thus, if an improved bound exists which applies also to random states, it must take into account the relative partition sizes in a many-body state    

%Thus, an improved bound needs to take  the relative partition sizes into account in a many-body state.

%%%%%%%%%%%%%%%%%%%%%%%%%%%%%%%%%%%%%%%%%%%%%%%%%%%%%%%%%%%%%%%%%%%%%%%%%%%%%%%%%%%%%%%%%%%%%%%%%%%%

\section{Conclusions}
\label{Conclu}

In this paper we have considered generalized GHZ and GHZ-diagonal states
evolving under the action of natural decoherence
processes. For these states we have derived analytical upper bounds to the
entanglement decay and shown that the fragility
of their entanglement increases exponentially as the size of the
system grows and independent of relative partition sizes. A comparison with random pure initial states under the same dynamics manifests the exponential fragility of entanglement towards
decoherence to be a distinct feature of GHZ-type states.

We stress once more that the class of entanglement quantifiers
considered in the calculations is very general. Thus, the present conclusions are not only valid for bipartite but also for any genuine
multipartite entanglement the system can present. Moreover, most
of the entanglement quantifiers aiming at the usefulness of a given quantum
state for certain applications that involve local operations fall into the category of quantifiers adopted in the present work~\cite{HorodeckiReview}. In this way, our results bound the GHZ-states' usefulness for most of quantum communicational and computational tasks.

\begin{acknowledgments}
We thank FAPERJ, the PROBRAL CAPES/DAAD Project,
The Brazilian Millenium Institute for Quantum Information,
EU QAP and COMPAS projects, Spanish MEC FIS2007-60182 and
Consolider-Ingenio QOIT projects and the Generalitat de Catalunya
for financial support. M. T., F. de M., and A. B. would like to express their gratitude for the hospitality of Luiz Davidovich's group in Rio de Janeiro. F. de M. acknowledges the financial support by Alexander von Humboldt foundation.
\end{acknowledgments}

%%%%%%%%%%%%%%%%%%%%%%%%%%%%%%%%%%%%%%%%%%%%%%%%%%%%%%%%%%%%%%%%%%%%%%%%%%%%%%%%
%
% Appendices
%
%%%%%%%%%%%%%%%%%%%%%%%%%%%%%%%%%%%%%%%%%%%%%%%%%%%%%%%%%%%%%%%%%%%%%%%%%%%%%%%%

\appendix

\section{Entangled contribution in the operator sum representation of depolarizing}
\label{app:Details(i)}

In the proof of \textit{(i)}, the application of the depolarizing
channel $\Lambda_\text{D}$ to the initial state
$\rho_0=\proj{\psi_0^+(\alpha,\beta)}$ yields, in the operator sum
representation using \eqref{DCPauli}, two contributions. The
application of operators $\sigma_1$ or $\sigma_2$ to any of the
qubits gives a separable contribution, whereas a
non-separable contribution can exclusively appear for the
application of only $\sigma_0=\openone$ or $\sigma_3$ to all the
qubits:
\begin{equation}
\sum_{j_1,\ldots,j_N=0,3} s_{j_1}\ldots s_{j_N} \sigma_{j_1}\otimes\cdots\otimes\sigma_{j_N} \rho_0 \sigma_{j_1}\otimes\cdots\otimes\sigma_{j_N}
\end{equation}
with respective prefactors $s_0=1-3p/4$ and $s_3=p/4$. An even number of parity changes due to $\sigma_3$ (and identity operators everywhere else) leaves the state effectively unchanged, $\proj{\psi_0^+(\alpha,\beta)}$, whereas an odd number changes the parity, $\proj{\psi_0^-(\alpha,\beta)}$.
The corresponding prefactors are then accordingly given by the sum over all even and odd powers of $s_3$ and $s_0$, respectively, weighted with their occurence:
\begin{align}
\lambda_+ &= \sum_{M=0, \text{even}}^N \binom{N}{M}
\left( 1-\frac{3p}{4}\right)^{N-M} \left(\frac{p}{4}\right)^M \\
\lambda_- &= \sum_{M=1, \text{odd}}^N \binom{N}{M}
\left( 1-\frac{3p}{4}\right)^{N-M} \left(\frac{p}{4}\right)^M
\,.
\end{align}
Their difference is what determines the contribution to the remaining coherences, by virtue of the binomial theorem:
\begin{align}
\lambda_+-\lambda_-
&=\sum_{M=0}^N \binom{N}{M} \left( 1-\frac{3p}{4}\right)^{N-M} \left(-\frac{p}{4}\right)^M
\\
&=(1-p)^N
\,.
\end{align}

%%%%%%%%%%%%%%%%%%%%%%%%%%%%%%%%%%%%%%%%%%%%%%%%%%%%%%%%%%%%%%%%%%%%%%%%%%%%%%%%%%%%%%%%%%%%%%%%%%%%%%%%%%%%%%%%%%%%%%%%%%%%%%%%

%%%%%%%%%%%%%%%%%%%%%%%%%%%%%%%%%%%%%%%%%%%%%%%%%%%%%%%%%%%%%%%%%%%%%%%%%%%%%%%%%%%%%%%%%%%%%%%%%%%%%%%%%%%%%%%%%%%%%%%%%%%%%%%%
%%%%%%%%%%%%%%%%%%%%%%%%%%%%%%%%%%%%%%%%%%%%%%%%%%%%%%%%%%%%%%%%%%%%%%%%%%%%%%%%%%%%%%%%%%%%%%%%%%%%%%%%%%%%%%%%%%%%%%%%%%%%%%%%

\begin{thebibliography}{99}
\bibitem{commentgray}
Notice that in all cases in Figs.~\ref{Fig1} and \ref{Fig2} the width of the distributions decreases with $N$, so that the very large majority of the negativities concentrates within a very thin region around their mean values~\cite{concentration}. This justifies the use of a (relatively) small sample of states as a valid representative to draw figures of merit for the whole Hilbert space; for additionally sampled states will most likely also fall within the thin region already accounted for.
\end{thebibliography}
\end{document}